\documentclass[12pt,a4paper]{article}

\usepackage{graphics}
\usepackage{latexsym}
\usepackage{amsmath}
\usepackage{amssymb}

\title{Trying to understand dark matter}
\author{B. Hoeneisen}
\date{\small{
Universidad San Francisco de Quito \\
3 February 2015} \\
}

\begin{document}
\maketitle

\begin{abstract}
\noindent
We present some ``back-of-the-envelope" calculations
to try to understand cold dark matter, 
its searches, and extensions of the 
Standard Model.
Some of the insights obtained from this exercise may be useful.
\end{abstract}

\section{Introduction}
For my own education I have done some ``back-of-the-envelope" calculations
to try to shed some light on these questions:
(i) What is cold dark matter (CDM) made of?
(ii) What are the most promising search strategies for CDM?
(iii) What are the simplest extensions of the Standard Model that
include CDM?
(iv) Why is the density of CDM not much greater than, or much less
than, the density of baryons?

\section{What do we know about cold dark matter?}
\label{know}

Cold dark matter is cold, i.e. non-relativistic at the time
of first galaxy formation, 
dark, i.e. interacts only very weakly with
electromagnetic radiation, and stable
on cosmological time scales, else it would have decayed by now.

The critical density of the universe is
\begin{equation}
\rho_c \equiv \frac{3 H_0^2}{8 \pi G_N} \equiv 1.878 \times 10^{-26} h^2 \textrm{ kg m}^{-3} 
= 1.054 \times 10^{-5} h^2 \textrm{ GeV cm}^{-3}
\end{equation}
with $h = 0.673 \pm 0.012$ \cite{PDG2014}.
The present density of baryonic matter in the universe
is $\rho_b \equiv \Omega_b \rho_c$ with \cite{PDG2014}
\begin{equation}
\Omega_b h^2 = 0.0221 \pm 0.0003.
\label{Omega_b}
\end{equation}
The present density of CDM in the universe
is $\rho_{\textrm{cdm}} \equiv \Omega_{\textrm{cdm}} \rho_c$ with \cite{PDG2014}
\begin{equation}
\Omega_{\textrm{cdm}} h^2 = 0.120 \pm 0.003.
\label{Omega_cdm}
\end{equation}

Structure formation in the Universe requires that the particles of CDM 
were non-relativistic at the onset of galaxy formation, i.e. at the
time when a galactic mass entered the horizon \cite{PDG2014}. 
A ``small" galaxy with baryonic plus dark mass $M \approx 10^8 M_\textrm{sun}$
entered the horizon at a photon temperature 
$T \approx 1.5 \times 10^7 \textrm{ K} = 1.3 \textrm{ KeV}$,
so particles of CDM have a mass $m_d > 1.3 \textrm{ KeV}$.
The co-moving velocity of non-relativistic non-interacting 
particles in an expanding Universe
varies as $\propto 1/a \propto T$, where $a$ is the expansion parameter. 
Therefore today, at $T = T_0 = 2.7255 \pm 0.0006$ K,
the velocity of particles of CDM referred to a homogeneous
Universe, i.e. before ``falling" into a galactic halo, is less
than $95 \textrm{ m/s}$.

We also require the
stronger condition $m_d > 1.3$ MeV so that particles of CDM
were non-relativistic at the time of Big Bang nucleosynthesis
in order to not upset the agreement between the predicted and
observed primordial abundances of $^4\textrm{He}$, D, $^3\textrm{He}$
and $^7\textrm{Li}$ \cite{PDG2014}.

If CDM particles interact with $Z$ with Standard Model coupling, 
then $m_d > \frac{1}{2} M_Z$ to not
upset the measured width of $Z$ decays.

The hierarchical formation of galaxies has been described elsewhere \cite{gal}.
Peaks in the density fluctuations in an expanding Universe grow and diverge acquiring
a density run $\rho \propto 1/r^2$, a mass inside radius $r$, $M(r) \propto r$, and
a velocity of circular orbits $\nu_0$ independent of $r$. 
These ``galactic halos" are composed of
CDM and baryonic matter. Baryonic matter interacts with 
baryons and photons,
radiates photons, and falls to the bottom of the halo potential well. This 
constitutes the visible matter of galaxies. The particles of the galactic
halo, including CDM and stars, have a root-mean-square velocity
\begin{equation}
\sqrt{\left< \nu^2 \right>} = \sqrt{\frac{3}{2}} \nu_0,
\label{vCDM}
\end{equation}
with the velocity of circular orbits $\nu_0$ typically 50 to 300 km/s.
The density of CDM is very inhomogeneous. For example, the local CDM
density is $2.4 \times 10^5$ times the mean. This inhomogeneity is due to
the hierarchical formation of galaxies \cite{gal}.

An important observation is that the ``bullet" galaxy cluster recently passed
through another cluster. The ordinary hot gas (composed of nucleons and electrons)
shocked and decelerated, while the CDM halos passed through each other
on ballistic trajectories \cite{PDG2014}. 
Therefore, CDM does not interact (or interacts
only very weakly) with nucleons, electrons, CDM and photons.

The only additional information we have on CDM are the negative
results of searches \cite{PDG2014} (some positive hints need
confirmation).

\section{Two scenarios}
\label{scenarios}
\subsection{Scenario I}
Let us assume that CDM is composed of
particles and antiparticles of mass $m_d$, with zero electric charge, and with 
equal number densities $\frac{1}{2} n_d$, that were once in
thermal equilibrium with the ultra-relativistic ``cosmological soup".
Note that, in Scenario I, we assume zero chemical potential.
In thermal equilibrium at temperature $T$, the density of ultra-relativistic
particles plus antiparticles is given by the Planck distribution
\begin{equation}
\rho_r = \frac{\pi^2}{30} \frac{(kT)^4}{\hbar^3 c^5} \left\{ N_b + \frac{7}{8} N_f \right\},
\label{rho_r}
\end{equation}
and the number density of 
particles plus antiparticles is
\begin{equation}
n_r = \frac{1.20205}{\pi^2} \left( \frac{kT}{\hbar c} \right)^3 \left\{ N_b + \frac{3}{4} N_f \right\}.
\label{n_rel}
\end{equation}
$N_b$ ($N_f$) is the number of boson (fermion)
degrees of freedom:
for electrons plus positrons $N_f = 4$, for photons $N_b = 2$. 
As the Universe cools, heavy particles become non-relativistic and
annihilate into lighter particles. These lighter particles 
heat up due to the annihilations.
The annihilation-creation reaction of CDM particles and
antiparticles $\nu_d$ and $\bar{\nu}_d$ can be written as
\begin{equation}
\nu_d + \bar{\nu}_d \leftrightarrow Y.
\label{annihilation}
\end{equation}
We assume that $Y$ is a set of
Standard Model particles (with mass less than $m_d$)
so that CDM, $Y$ and photons
remain in thermal equilibrium until the freeze-out
temperature $T_f$. 
(The case with $Y$ \textit{not} a set of Standard Model
particles will be considered later.)
The number density of non-relativistic particles plus antiparticles,
in thermal equilibrium at temperature $T$, with zero chemical potential, is
\begin{equation}
n_d = N_{f, b} \left( \frac{m_d k T}{2 \pi \hbar^2} \right)^\frac{3}{2}
\exp{ \left[ - \frac{m_d c^2}{k T} \right] },
\label{nd_non_rel}
\end{equation}
(in the scenario being considered, we take $N_f = 4$).
Thereafter, the number density of CDM particles and antiparticles
decreases in proportion	to $a^{-3}$, where $a$ is the
expansion parameter of the Universe.                           

Freeze-out occurs when the mean time to annihilation of a CDM particle
\begin{equation}
\tau = \frac{2}{n_d \sigma v_d}
\label{tau}
\end{equation}
exceeds the age of the radiation dominated Universe
\begin{equation}
t = \frac{1}{2} \sqrt{ \frac{3}{8 \pi G_N \rho_r} }.
\label{t}
\end{equation}

Entropy is conserved while annihilation-creation reactions remain in equilibrium.
The entropy density of ultra-relativistic particles is
\begin{equation}
s \equiv \frac{S}{V} = \frac{2 \pi^2}{45} \left( \frac{kT}{\hbar c} \right)^3 
\left( N_b + \frac{7}{8} N_f \right).
\end{equation}
As particles become non-relativistic and annihilate,
$a^3 T^3 \left( N_b + \frac{7}{8} N_f \right)$
of the ultra-relativistic particles remains constant.
The effective value of $\left( N_b + \frac{7}{8} N_f \right)$ today
is 3.36 (it includes photons and the $\approx 30 \%$ cooler neutrinos).
In Table \ref{CDM} we set 
$\left( N_b + \frac{7}{8} N_f \right) = 3.36, 10.75$ and $86.25$
at freeze-out for $m_d = 10^{-3}, 10^0$ and $10^3$ GeV 
respectively \cite{PDG2014}. 
As an estimate, for $m_d = 10^6$ GeV we take 
$N_b = 2 + 3 \cdot 3$ for $\gamma$, $W^+$, $W^-$ and $Z$,
and $N_f = 4 \cdot (6 \cdot 3 + 3 + 1.5)$ for six quarks plus three
charged leptons plus three neutrinos, 
so $\left( N_b + \frac{7}{8} N_f \right) \approx 89.75$
at freeze-out.

Today, the number density of particles plus antiparticles of CDM is
\begin{equation}
n_{d0} = \frac{\Omega_\textrm{cdm} \rho_c}{m_d}.
\label{nd0}
\end{equation}
At freeze-out,
\begin{equation}
n_d = n_{d0} \left( \frac{a_0}{a_f} \right)^3
= n_{d0}\left( \frac{T_f}{T_0} \right)^3 \frac{N_b + \frac{7}{8} N_f}{3.36}.
\label{nd}
\end{equation}

We solve these equations iteratively as follows. Due to the
exponential in Eq. (\ref{nd_non_rel}), the value of
\begin{equation}
\kappa \equiv \frac{m_d c^2}{k T_f}
\label{kappa}
\end{equation}
is close to 20. With this initial value of $\kappa$, and for each given $m_d$,
we obtain $T_f$ from Eq. (\ref{kappa}), and $v_d \approx \sqrt{\frac{3}{\kappa}} c$.
We obtain $n_d$ at freeze-out with Eq. (\ref{nd_non_rel}), and again with Eq.
(\ref{nd}), and
adjust $\kappa$ until both calculations of $n_d$ agree.
Finally we obtain the age $t_f$ of the Universe at freeze-out
from Eqs. (\ref{rho_r}) and (\ref{t}), and the CDM 
particle-antiparticle annihilation cross-section $\sigma$ from
Eq. (\ref{tau}) with $\tau = t_f$.

From these equations we obtain the results presented in Table \ref{CDM} 
for each $m_d$. It is interesting to note that the annihilation 
cross-section, needed to obtain the observed CDM density,
is of the order expected for the weak interaction, and is
quite insensitive to $m_d$. Note that most CDM particles have
annihilated, i.e. $n_d/n_\gamma \ll 1$.

\begin{table}
\begin{center}
\begin{tabular}{|cccccc|}
\hline
$m_d$ & $n_d/n_\gamma$ & $\kappa$ & $T_f$ & $t_f$ & $\sigma$ \\
$[$GeV$]$ & today &  & $[K]$ & $[$s$]$ & $[$pb$]$ \\
\hline
$1 \times 10^{-3}$ & $3.1 \times 10^{-6}$ & $16.99$ & $6.8 \times 10^8$ & $3.8 \times 10^{+2}$ &
$21.0$ \\
$1 \times 10^{+0}$  & $3.1 \times 10^{-9}$ & $23.20$ & $5.0 \times 10^{11}$ & $4.0 \times 10^{-4}$ &
$18.7$ \\
$1 \times 10^{+3}$  & $3.1 \times 10^{-12}$ & $28.32$ & $4.1 \times 10^{14}$ & $2.1 \times 10^{-10}$ &
$8.9$ \\
$1 \times 10^{+6}$  & $3.1 \times 10^{-15}$ & $35.53$ & $3.3 \times 10^{17}$ & $3.2 \times 10^{-16}$ &
$12.3$ \\
\hline
\end{tabular}
\end{center}
\caption{
For each CDM particle mass $m_d$ we calculate, for Scenario I,
the present number density of CDM particles plus antiparticles $n_d$
relative to the photon number density $n_\gamma$,
$\kappa \equiv m_d c^2 / (k T_f)$,
the photon temperature at freeze-out $T_f$, the age of the Universe at
freeze-out $t_f$, and the center of mass
annihilation cross-section $\sigma$ at freeze-out needed to obtain
the present CDM density.
}
\label{CDM}
\end{table}

\subsection{Scenario II}
The present CDM density $\rho_\textrm{cdm}$ is determined either by the annihilation cross-section
if annihilation is not complete (Scenario I), 
or by an asymmetry of CDM particles and antiparticles if 
annihilation is complete (Scenario II).
In Scenario II we assume a primordial asymmetry
$(n_d - \bar{n}_d)/n_\gamma$ equal to the present ratio $n_d/n_\gamma$
given in Table \ref{CDM}, and an annihilation cross-section 
$\sigma$ greater than given in Table \ref{CDM} so that
annihilation is complete.

From observations, $5 \Omega_b \approx \Omega_\textrm{cdm}$.
The fact that $\Omega_b$ is not much less than, or much greater than
$\Omega_\textrm{cdm}$, suggests that the values of $\Omega_b$ and
$\Omega_\textrm{cdm}$ are related.
If Scenario II is correct, then the particle-antiparticle
differences relative to $\gamma$'s, for ultra-relativistic CDM and baryons, 
are in the approximate ratio $5 m_p/m_d$.

\section{Bench mark annihilation cross-sections}
\label{neutrino}
(Note: in this Section we omit the symbols $c$ and $\hbar$.)
Let us consider a heavy Dirac particle $\nu_d$ with
mass $m_d > \frac{1}{2} M_Z$, with the Standard Model neutrino coupling to $Z$.
We calculate the annihilation cross-section for
$\nu_d \bar{\nu}_d \rightarrow Z^* \rightarrow e^+ e^-$
for non-relativistic $\nu_d$, and ultra-relativistic $e^+$ and $e^-$. 
In the center of mass, we obtain
(after summing over final state polarizations, and averaging over
initial state polarizations):
\begin{equation}
\sigma = \frac{1}{\pi v_d} \left( \frac{e}{4 \sin{\theta_W} \cos{\theta_W}} \right)^4
\frac{m_d^2}{(4 m_d^2 - M_Z^2)^2} 
\left( 1 - 4 \sin^2 \theta_W + 8 \sin^4 \theta_W \right),
\label{sigma_cdm}
\end{equation}
where $v_d$ is the velocity of the incident particles in the center of mass. 
At freeze-out the velocity of CDM particles is $v_d \approx \sqrt{3/\kappa} c \approx 0.35 c$,
so the non-relativistic equation is an approximation.
Including final states with $\nu_e$, $\nu_\mu$, $\nu_\tau$, $e$, $\mu$, $\tau$,
$u$, $c$, $t$, $d$, $s$ and $b$ we obtain
\begin{equation}
\sigma = \frac{8}{\pi v_d} \left( \frac{e}{4 \sin{\theta_W} \cos{\theta_W}} \right)^4
\frac{m_d^2}{(4 m_d^2 - M_Z^2)^2}
\left( 3 - 6 \sin^2 \theta_W + 8 \sin^4 \theta_W \right), 
\label{sigma_cdm2}
\end{equation}
(we have neglected 
the $\gamma \gamma$ final state for this estimate).
The problem is then to find the mass $m_d$ that obtains the annihilation
cross-section corresponding to the observed dark matter density.
The result is $m_d \approx 205$ GeV and $\sigma \approx 10$ pb, for Scenario I. 
It is amazing that the weak interaction produces $\Omega_\textrm{cdm}$
of the correct order of magnitude!

For comparison, we consider the electromagnetic annihilation 
$e^+ e^- \rightarrow \gamma \gamma$ with
non-relativistic electrons of mass $m_e$. 
We obtain an annihilation cross-section
\begin{equation}
\sigma = \frac{e^4}{16 \pi v_e m_e^2}.
\end{equation}
At the freeze-out velocity $v_e \approx 0.35 c$ this cross-section is 
$\sigma = 7 \times 10^{11}$ pb, so the electromagnetic
annihilation of $e^+ e^-$ is complete.

In summary, a weakly interacting heavy particle $\nu_d$ (WIMP), with zero 
electric charge, and the Standard Model couplings of a neutrino to $Z$, $W^+$ and $W^-$,
can have the correct annihilation cross-section to produce the
observed $\Omega_\textrm{cdm}$ in Scenario I. Even Scenario II 
is possible. 
We note that a sterile neutrino, with a coupling to $Z$ weaker than in the
Standard Model, will produce an $\Omega_\textrm{cdm}$ 
that is larger than observed, unless there is an enhancement of
the annihilation cross-section due to a resonance with $m_d$ slightly
greater than $\frac{1}{2}M_Z$.
A stable composite neutral particle with a typical hadron cross-section 
(a few mb) is ruled out.

Finally, for future use, we calculate the elastic scattering 
cross-section of $\nu_d e^- \rightarrow \nu_d e^-$
with exchange of a virtual $Z$, for non-relativistic $\nu_d$ and $e^-$.
We obtain
\begin{equation}
\sigma = \frac{e^4 m_d^2 m_e^2 ( 1 - 2 \sin^2{\theta_W} + 4 \sin^4{\theta_W} )}
{64 \cdot \pi ( m_d + m_e )^2 M_Z^4 \cdot \sin^4{\theta_W} \cdot \cos^4{\theta_W}}.
\label{scattering}
\end{equation}

\section{Search strategies for dark matter}
\label{bench}
\subsection{Indirect detection}
We consider the search for narrow $\nu$ resonances from 
CDM annihilations
$\nu_d \bar{\nu}_d \rightarrow Z^* \rightarrow \nu \bar{\nu}$.
(These arguments also apply to the final states $e^+ e^-$ and $p^+ p^-$.)
Neutrinos from CDM annihilation have an energy $m_d$ with a relative half width 
approximately $2 v_d / c \approx 0.1 \%$. 
The number $F$ of $\nu$ plus $\bar{\nu}$ that a telescope receives, per second, ster-radian and
m$^2$ of telescope area, from CDM particle-antiparticle annihilations
within a distance $l$ is
\begin{equation}
F = \frac{1}{8 \pi} \left< n_d^2 \right> \sigma v_d l,
\end{equation}
where $n_d$ is the present number density of
CDM particles plus antiparticles (it is a function of the
position in the galactic halos), $\sigma$ is their annihilation cross-section,
and $v_d$ is their mean velocity. 
$n_\nu$ is the present number density of neutrinos plus antineutrinos in the Universe
from CDM annihilation:
\begin{equation}
n_\nu = \frac{1}{2 c} \left< n_d^2 \right> \sigma v_d l.
\label{nu}
\end{equation}
For this estimate we use the following
bench mark numbers: $\sigma = 10$ pb (see Table \ref{CDM}), 
$v_d = 100$ km/s (since most dark matter
particles have fallen into galaxy halos) \cite{gal},  
and we take $l$ to be the distance to the Universe horizon  
(corresponding to the case of no neutrino absorption). 
An important point is that the spatial mean $\left< n_d^2 \right>$ is
much larger than $\left< n_d \right>^2$. We take
$\left< n_d^2 \right> \equiv \alpha \left< n_d \right>^2$, with
$\alpha$ of order the ratio of the local CDM density in the 
Galaxy ($\approx 0.3$ GeV$/$cm$^3$) to the mean CDM density of the Universe
($\Omega_\textrm{cdm} \cdot \rho_c = 1.26 \cdot 10^{-6}$ GeV$/$cm$^3$),
i.e $\alpha \approx 2.4 \times 10^5$.
The resulting order-of-magnitude estimate is
\begin{equation}
F \approx \left( \frac{100 \textrm{GeV}}{m_d} \right)^2 \cdot 6 \times 10^{-8}
\left[ \frac{\textrm{1}}{\textrm{s} \cdot \textrm{sr} \cdot \textrm{m}^2} \right] .
\label{F2}
\end{equation}

Let us now calculate the number of interactions of $\nu$ plus $\bar{\nu}$
due to CDM annihilation per kilogram of detector material and per second:
\begin{equation}
R = \frac{n_\nu c A \sigma_N}{m_p} \approx \left( \frac{A}{28} \right) \cdot
\left( \frac{100 \textrm{ GeV}}{m_d} \right) \cdot
6 \times 10^{-19}
\left[ \frac{\textrm{interactions}}{\textrm{kg} \cdot \textrm{s}} \right].
\label{R}
\end{equation}
$n_\nu$ is the number density of $\nu$ plus $\bar{\nu}$ in the Universe
due to CDM annihilation, $E_\nu = m_d$ is their
average	energy in the laboratory frame,
$\sigma_N = E_\nu \times 0.005$ pb/GeV is their
average cross section with a nucleon \cite{PDG2014}, $A$ is
the number of nucleons (protons plus neutrons) in the nucleus of
the detector material atoms, and $m_p$ is the nucleon mass.

From Eq. (\ref{R}) we conclude that
the observation of monochromatic neutrinos from CDM annihilation
is out of experimental reach.
From Eq. (\ref{F2}) we conclude that the observation 
of monochromatic positrons or anti-protons from CDM annihilation
may be possible depending on the background levels and absorption.

\subsection{Direct detection}
\label{limits}
We consider elastic collisions 
$\nu_d q \rightarrow Z^* \rightarrow \nu_d q$,
where $q$ is a nucleus with $A$ nucleons and mass $A m_p$. 
The cross-section for this collision is $\sigma_q$.
The number of collisions of $\nu_d$ and $\bar{\nu}_d$ 
per kilogram of target material per second is
\begin{equation}
R = \frac{\rho_\textrm{cdm} \sigma_q v_d}{m_d A m_p}.
\end{equation}
We take the following bench mark numbers: local density of CDM
$\rho_\textrm{cdm} = 0.3$ GeV/cm$^3$; velocity of CDM particles 
in the laboratory frame $v_d = 100$ km/s;
mass of CDM particles $m_d = 100$ GeV, and take silicon 
(with $A = 28$) as a target nucleus. 

We \textit{estimate} $\sigma_q$ 
from the cross-section (\ref{scattering}) of the
reaction $\nu_d e^- \rightarrow \nu_d e^-$, by replacing $m_e$
with $A m_q$ for coherent scattering
on $A$ nucleons with an effective mass $m_q$.
To \textit{estimate} $m_q$ we compare the cross-section
\begin{equation}
\sigma = \frac{G_F^2 s}{\pi} = \frac{G_F^2}{\pi} 2 E_\nu m_q
\end{equation}
of reaction $\nu_\mu d \rightarrow \mu^- u$, with the experimental
cross-section for quasi-elastic scattering $\nu_\mu n \rightarrow \mu^- p$:
$\approx 0.012$ pb/GeV $\times E_\nu$ for $E_\nu < 1$ GeV \cite{PDG2014}. 
$E_\nu$ is the laboratory
energy of the neutrino. We obtain $m_q \approx 0.36$ GeV, and
$\sigma_q \approx (A/28)^2 \cdot 0.56 \textrm{ pb} = A^2 \cdot 7 \times 10^{-40}$ cm$^2$.
Finally
\begin{equation}
R \approx \left( \frac{A}{28} \right) \cdot
\left( \frac{100 \textrm{ GeV}}{m_d} \right) \cdot 
\left( \frac{v_d}{100 \textrm{km/s}} \right)
\cdot 3.6 \times 10^{-7}
\left[ \frac{\textrm{interactions}}{\textrm{kg} \cdot \textrm{s}} \right]. 
\end{equation}
In conclusion, the observation of CDM-nucleon elastic scattering is within the reach of
direct detection experiments if the cross-section is comparable
to that	of a heavy neutrino with Standard Model	coupling to $Z$.

\subsection{Collider experiments}
Dark matter may be produced at colliders. For example,
$u \bar{u} \rightarrow \gamma Z^*$ with
$Z^* \rightarrow \nu_d \bar{\nu}_d$.
The event selection requires a single $\gamma$ (or single
jet) with high transverse momentum $p_T$, and high
back-to-back missing transverse energy (since dark matter leaves
no trace in the detector). 
This dark matter ``signal" must compete with the Standard Model
background $Z^* \rightarrow \nu \bar{\nu}$, and so can be
observed only if the coupling of $Z$ to dark
matter is larger than the Standard Model coupling of $Z$ to $\nu$.
The limits obtained \cite{cdm_LHC} are of the order of the
cross section calculated above for a heavy neutrino with the
Standard Model coupling to $Z$, i.e.
$\approx 7 \times 10^{-40}$ cm$^2$ per nucleon.

\section{Direct searches}
\label{direct}
The LUX experiment \cite{LUX} is a dual-phase xenon time-projection chamber
that measures nuclear recoils.
The LUX Collaboration has determined that the
CDM-nucleon elastic cross-section is less than
$10^{-44}$ cm$^2$ to $10^{-45}$ cm$^2$ for $m_d$ in the range
$10$ GeV to $10^3$ GeV. A similar limit has been 
set by the XENON100 experiment.
In comparison, the CDM-nucleon elastic cross-section
for the Standard Model
$\nu$-$Z$ coupling was calculated (in Subsection \ref{limits})
to be $\approx 7 \times 10^{-40}$ cm$^2$ independent of $m_d$.

In conclusion, a CDM particle with the Standard Model
$\nu$-$Z$ coupling has been ruled out by the direct
search experiments if $m_d$ is in the range $10$ GeV to $10^3$ GeV.
If $m_d > 10^3$ GeV, a CDM-$Z$ coupling \textit{greater} than
the Standard Model $\nu$-$Z$ coupling is needed to reach
Scenario I. If $m_d < 10$ GeV, a CDM-$Z$ coupling \textit{less} than
the Standard Model $\nu$-$Z$ coupling is needed to not
upset the $Z$ width, and then Scenario I can not be
reached. If CDM couples weakly to $Z$, so
the CDM-nucleon elastic cross-section is at the LUX limit of
$10^{-44}$ cm$^2$, then already some fine tuning is necessary to
reach Scenario I: $m_d$ must be within 0.7 GeV of $\frac{1}{2}M_Z$,
see the denominator in Eq. (\ref{sigma_cdm2}).

\begin{figure}
\begin{center}
\vspace*{-0.7cm}
\scalebox{0.6}
{\includegraphics{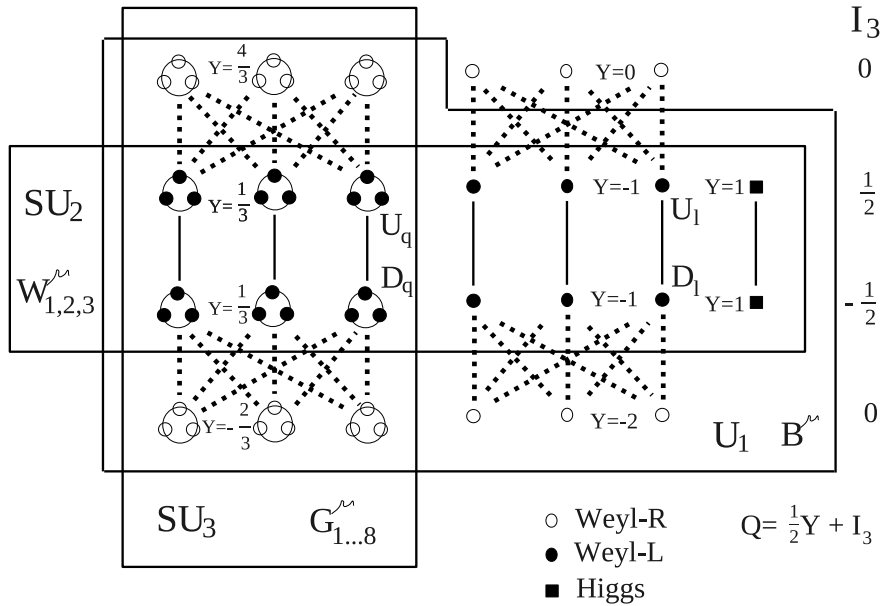}}
\vspace*{-8.5cm}
\caption{
The elementary fields of the Standard Model
and their local symmetries $U_1 \times SU_2 \times SU_3$.
$SU_2$ transformations are indicated by solid vertical lines.
$SU_3$ transformations are indicated by circles with three
black or three white dots. How can this diagram be extended to
include a candidate of CDM?
}
\label{SM}
\end{center}
\end{figure}

\section{Extensions of the Standard Model}

\subsection{The Standard Model}
The elementary fields of the Standard Model plus three right-handed
neutrinos are presented in Figure \ref{SM}. This beautiful
``cathedral" is solid (it has many ``cross-beams") and can not be
modified ``a little bit" because it is based on symmetries:
12 global $SU_3$ symmetries, indicated by circles that mix
three Weyl-L fields (black dots) or three Weyl-R fields (white dots);
12 global $SU_2$ symmetries, indicated by vertical bars that mix
two Weyl-L fields, and global $U_1$ symmetries for all dots
except the right handed neutrino fields. These Weyl-L and Weyl-R
fields are the two inequivalent irreducible 2-dimensional representations
of the group of proper Lorentz transformations. So far, these fields
are massless and have no interactions. Now, one common $SU_3$ symmetry,
and one common $SU_2$ symmetry, and one common $U_1$ phase symmetry 
are promoted to be \textit{local} symmetries. To this end it is necessary
to replace the ordinary derivatives by \textit{covariant} derivatives
which contain ``connectors" (called ``gauge" fields),
one for each generator of the local symmetry groups:
8 ``gluons" $G_i^\mu$ for $SU_3$, $W_1^\mu$, $W_2^\mu$ and $W_3^\mu$ 
for $SU_2$, and $B^\mu$ for $U_1$.
The fields are functions of only \textit{one} space-time point. 
The ordinary derivatives compare the fields at two
neighboring space-time points (or two neighboring ``lattice points" 
in the lattice approximation). Hence the ``connectors" 
depend on \textit{two} neighboring ``lattice" points in space-time 
in order to obtain \textit{local} symmetries (that is why the ``connectors"
have ``gauge transformations" due to the local symmetry transformations that
are different at the two points) and cause
the interactions between the fields.
So far, all fields and connectors are massless. Finally, a 
complex Higgs doublet $\Phi$ with respect to the local $SU_2$ symmetry, 
and singlet with respect to the local $SU_3$ symmetry,
with a $U_1$ quantum number $Y = 1$, is added to the model, with a self-potential
that gives it a vacuum expectation value $v$ at low temperature. Several
``miracles" occur. The $W_1^\mu$, and $W_2^\mu$, and a linear
superposition of $W_3^\mu$ and $B^\mu$, acquire 
a third (longitudinal) amplitude needed to become massive (``swallowing" up three of the
four amplitudes of the Higgs field), and all particles except the photon,
acquire mass due to $v$ in such a way that the theory remains ``renormalizable".
Furthermore the Weyl-L and Weyl-R fields become ``tied together" into
massive ``Dirac" fields as indicated by the dotted lines in Figure \ref{SM}.
Finally, each ``family" of two quarks and two leptons cancel ``triangle anomalies",
so no dot in Figure \ref{SM} can be removed (except one of the $\nu_R$). 
Experimental limits on
``flavor changing neutral currents" severely limit the allowed extensions
of the Standard Model.

Note that the electron is not an elementary field: it is composed of
a Weyl-R field and a superposition of three Weyl-L fields that
forward scatter on $v$, thereby acquiring mass \cite{mass}.
The scattering is forward because $v$ does not depend on the space-time
point $(t, x, y, z)$. These scatterings change a Weyl-L field into a Weyl-R
field, and vice-versa, resulting in a massive Dirac field.
Even the $\gamma$ and $Z$ are
not elementary fields: they are two orthogonal superpositions of $W_3^\mu$ and $B^\mu$.
Finally, the ``quantum collapse" of the fields produce 
\textit{particles} on their mass shell, 
in accordance with the Planck and De Broglie relations \cite{mass}.

%We define fields according to their irreducible representations
%with respect to the proper Lorentz group, and groups $SU_2$ and $SU_3$, and
%the charge $Y$ with respect to U(1). 
%The Higgs field $\Phi$ is (scalar, 2, 1, $Y=1$),
%and its conjugate $\Phi_c = i \tau_2 h^*$ is (scalar, 2, 1, $Y=-1$).
%In a special gauge
%\begin{equation}
%\Phi = \frac{1}{\sqrt{2}} \left(
%\begin{array}{c}
%0 \\
%v + h
%\end{array} \right)
%, \qquad
%\Phi_c = \frac{1}{\sqrt{2}} \left(
%\begin{array}{c}
%v + h^* \\
%0
%\end{array} \right),
%\label{h}
%\end{equation}
%where $v$ is the vacuum expectation value of the Higgs field, and $h$ is
%the only remaining amplitude of the Higgs field in this gauge.

\subsection{The Standard Model is incomplete}
The Standard Model, with all of its successes over 23 orders of magnitude in
energy (from the Lamb shift in hydrogen to the experiments at the LHC), 
does not explain three observed phenomena: 
(i) neutrino oscillations,
(ii) the baryon asymmetry of the Universe, and
(iii) dark matter.
We can not change the Standard Model a ``little bit", but we can add new
fields and local symmetries.
We have already added three Weyl-R fields to Figure \ref{SM}
and ``Yukawa mass terms" to the neutrinos to allow
neutrino oscillations. 

In what direction should we extend the Standard Model to
incorporate CDM that does not interact significantly with baryons, CDM or photons? 
The LHC experiments do not see any signal
beyond the Standard Model in the data of the runs at $\sqrt{s} = 7$ or $8$ GeV.
Precision measurements at LHC, Tevatron, Babar, Belle, and elsewhere,
reveal no new-physics discovery with $5 \sigma$ confidence.
At the present time there is no experimental result that requires
new local symmetries, i.e. new interactions: no $Z'$ or $W'$ has been found at the 
Tevatron or LHC.
Furthermore, as soon as we add a second Higgs doublet  
(as in the 2-Higgs doublet models, or in supersymmetry), 
new global symmetries are needed
to avoid ``flavor changing neutral currents" \cite{Branco}.
The measured mass of the Higgs $m_h = 125.7 \pm 0.4$ GeV \cite{PDG2014} is very special: 
at this mass the Standard Model is valid and calculable, i.e. perturbative, all the
way up to the Planck energy \cite{PDG2014}, so there is no experimental need
for a grand unified theory (GUT) with its ``hierarchy problem",
and hence no experimental need for supersymmetry (SUSY) or technicolor 
to solve the hierarchy problem \cite{PDG2014}.
If the Higgs mass were greater than observed, the Higgs self-coupling
would become non-perturbative at some scale \cite{PDG2014}.
The observed Higgs mass is consistent with the upper bound
from perturbative unitarity constraints \cite{PDG2014}.
A low value of the Higgs mass would have favored a two Higgs doublet
model, or a SUSY extension of the Standard Model.
The measured Higgs mass is special: at this mass particles acquire mass
by the ``stepping stone" process, not the ``particle in a box" model \cite{mass}.

Before the LHC, we knew that new physics was right around the corner.
Now we do not see the corner \cite{Illiopoulos}.

In the following Subsections we consider extensions of the Standard
Model with CDM particles and antiparticles annihilating
into (i) Standard Model particles, or (ii) non-Standard Model particles.
Cases (i) are more predictive because CDM was once in thermal equilibrium with
the ``cosmological soup". CDM particle-antiparticle annihilation
in cases (i) heat up the Standard Model particles and hence do not
change the predictions of Big Bang nucleosynthesis. Cases (ii) are 
still allowed by nucleosynthesis if there is no more than one
nearly massless ``dark" neutrino into which CDM can decay \cite{PDG2014}.

\subsection{Heavy neutrino}
The Standard Model has no CDM candidate, so we must add
dots (i.e. fields), and perhaps local symmetries, to Figure \ref{SM}.
Let us consider the addition of an $SU_2$ multiplet of
dots, to the area labeled ``$SU_2$" in 
Figure \ref{SM}. Once we choose the multiplet, e.g. a \textbf{2} or \textbf{3},
we need to choose the spin of these fields, e.g. 
0 (scalar), $\frac{1}{2}$ (Weyl-L or Weyl-R), or 1 (vector).
Then we must add mass terms to the lagrangian
of the form $- m^2 \chi \chi$ or $-G \Phi^\dagger \Phi \chi \chi$ for
spin 0 or spin 1. For spin $\frac{1}{2}$ we can choose the
Standard Model form
$- G_u [ \widetilde{\chi}_\textrm{2L} \Phi_c \chi_\textrm{1Ru} +
\widetilde{\chi}_\textrm{1Ru} \Phi^\dagger_c \chi_\textrm{2L} ]
- G_d [ \widetilde{\chi}_\textrm{2L} \Phi \chi_\textrm{1Rd} +
\widetilde{\chi}_\textrm{1Rd} \Phi^\dagger \chi_\textrm{2L} ]$, 
or $L \leftrightarrow R$, so we need to
add the corresponding singlet dots outside of the ``$SU_2$" area
in Figure \ref{SM}. 

We note that the CDM $SU_2$ coupling is the same as in the Standard
Model: it can not be ``turned down" because the $SU_2$ group is
non-abelian. Therefore,
\textit{all of these models have been ruled out} by the 
nucleus recoil searches as discussed in Section \ref{direct}.

\subsection{Sterile neutrino}
To the Standard	Model lagrangian
we have already added three Weyl-R fields
with Yukawa mass terms, see Figure \ref{SM}. This extension adds mass to the
massless Standard Model neutrinos, and
describes neutrino oscillations. If the	neutrinos
are Majorana particles,	i.e. if	the neutrinos are
identical to their anti-neutrinos, then it is possible, in addition
to the Yukawa mass terms,
to add Majorana mass terms to the $\nu_R$ (which have $Y = 0$).
Below the energy scale of $M_W$
we can replace the Higgs field by its vacuum expectation
value. Then the lagrangian takes the form \cite{nuSM}
\begin{eqnarray}
\mathcal{L} & = & \mathcal{L}_{SM} 
+ \frac{i}{2} \tilde{\nu}_R \gamma^\mu \partial_\mu \nu_R -
\frac{i}{2} \partial_\mu \tilde{\nu}_R \cdot \gamma^\mu \nu_R - \nonumber \\
& & m_D \widetilde{\nu}_L \nu_R - m^*_D \widetilde{\nu}_R \nu_L
- \frac{1}{2} M \widetilde{\nu^c}_R \nu_R - \frac{1}{2} M^* \widetilde{\nu}_R \nu^c_R.
\end{eqnarray}
We have suppressed family indices.
$\mathcal{L}_{SM}$ is the Standard Model lagrangian.
$m_D$ is a Yukawa mass matrix. $M$ is a Majorana mass matrix.
For simplicity, we consider a single family.
We take $|m_D| \ll M$.
Diagonalizing the mass terms we obtain, to first order in ${|m_D|}/{M}$,
\begin{eqnarray}
\nu_\textrm{active} & = & \nu_L - \frac{m_D}{M} \nu_R \textrm{   with mass }  \frac{|m_D|^2}{M}, \nonumber \\
\nu_\textrm{steril} & = & \nu_R + \frac{m^*_D}{M} \nu_L \textrm{   with mass }  M.
\end{eqnarray}
For CDM, $M > 1.3$ MeV, see Section \ref{know}.
Then, for an active neutrino mass $\approx 0.05$ eV, we obtain $|m_D|/M < 2 \times 10^{-4}$.
Note that the steril neutrino has a small $\nu_L$ component that interacts with $Z$.
To obtain sufficient dark matter annihilation it is necessary to be near resonance
in Eq. (\ref{sigma_cdm2}), i.e. 
$m_d - \frac{1}{2}M_Z$ must be less than $0.9 \times 10^{-7} \cdot \frac{1}{2}M_Z$.
This result illustrates the degree of fine tuning that steril neutrino CDM requires.
The steril neutrinos may solve the baryon asymmetry of the Universe in addition
to the dark matter problem \cite{nuSM}.

\subsection{$Z'$}
Let us consider the addition of a new local symmetry to Figure \ref{SM},
and add a multiplet of new dots with this symmetry. 
Some Standard Model particles must also form a multiplet of the new symmetry.
As an example, the new local symmetry could be
a new $SU_2$, which we call $SU'_2$, and the new multiplet could be
a Weyl-R doublet. Then we must add a complex Higgs field that is a
$SU'_2$ doublet, and two Weyl-L singlets to complete the massive
Dirac (Weyl-L $\oplus$ Weyl-R) fields. 
Note that we have chosen Weyl-R doublets for the CDM, contrary to
the Weyl-L doublets of the Standard Model, to avoid their mixing and
obtain a stable CDM particle. In this particular example we
include (some?) Standard Model leptons, but not the quarks, in the
local $SU'_2$ symmetry, to evade the nucleus recoil experiments.
Coupling to quarks is allowed if $m_d < 10$ GeV, or if the coupling
of $SU'_2$ is less than the coupling of $SU_2$ and annihilation is
near resonance.
These models contain new gauge connectors, e.g. $W'_1$, $W'_2$, $W'_3$, 
new particles, and a new Higgs boson which can be searched at the LHC,
or at a future $e^- e^+$ linear collider.

\subsection{Dark world}
\label{dw}
Let us consider dark matter with \textit{no} interactions with
Standard Model particles. Therefore we consider two separate sectors
with no common local symmetry, i.e. no mutual interactions except gravity:
the Standard Model ``cathedral" of Figure \ref{SM}, and a ``dark world". 
The equations of Section \ref{scenarios} are valid separately
for each sector. The two sectors have in common the same Universe expansion
factor $a$, and the same space-time geometry,
but may have different temperatures $T$ and $T'$ 
(in this Section the prime denotes parameters of the dark world, and the subscript 0
denotes present day values).
Entropy is conserved separately in both sectors, so
\begin{equation}
\frac{T}{T_0} = \frac{a_0}{a} \left[ \frac{g_0}{g} \right]^\frac{1}{3},
\qquad
\frac{T'}{T'_0} = \frac{a_0}{a} \left[ \frac{g'_0}{g'} \right]^\frac{1}{3},
\end{equation}
where $g \equiv N_b + \frac{7}{8}N_f$.

Consider a ``dark world" with the same ``cathedral" of Figure \ref{SM}.
Such dark matter would collide with itself radiate dark photons and
fall to the bottom of the galactic halos. So we turn off the strong interaction by not requiring
the $SU'_3$ symmetry to be local. The dark photons, in addition to 
the $\nu'$, would upset Big-Bang
nucleosynthesis, so we also do not require the $U'_1$ symmetry to be local.
The only remaining connectors, in this example, are $W'_1$, $W'_2$ and $W'_3$.
We need at least two families. The lightest particle of family 2 is
the CDM candidate, and the lightest particle of family 1 is a light
neutrino, which is the only allowed particle into which CDM can annihilate
and still be compatible with Big Bang nucleosynthesis.
However there are few
experimental constraints, so, for the present, the ``dark world" is
not very predictive.

\section{Conclusions}
\label{conclusion}
\begin{itemize}
\item
CDM with the Standard
Model coupling of neutrinos to $Z$ is ruled
out by nucleus recoil searches and the measured $Z$ width.
\item
CDM composed of steril neutrinos is possible,
and may also be compatible with the baryon asymmetry of the Universe,
but needs fine tuning.
Experiment: Push the limit on neutrinoless double beta decay;
sattelite searches for monochromatic $e^+$ and $p^-$ with energy near $\frac{1}{2} M_Z$.
\item
CDM may have new interactions with some Standard Model particles.
Experiments: Search for $Z'$, $W'$, new particles, and new Higgs boson, 
at the LHC, and at a future $e^- e^+$ linear collider. 
Further discussion is presented below.
\item
CDM may be part of a ``dark world", with no interactions
with Standard Model particles except gravity.
Experiment: Tighten constraints from Big Bang nucleosynthesis.
\end{itemize}

Why is the density of CDM not much greater than, or much less
than, the density of baryons? The observations 
$\Omega_\textrm{cdm} \approx 5 \Omega_b$, $n_b/n_\gamma \ll 1$ and
$n_\textrm{cdm}/n_\gamma \ll 1$, suggest that there is
a common origin of $\Omega_\textrm{cdm}$ and $\Omega_b$.
The source of $\Omega_b$ is a primordial baryon asymmetry of the
Universe. $\Omega_\textrm{cdm}$ and $\Omega_b$ \textit{may be related} if
(i) $\Omega_\textrm{cdm}$ is due to a primordial asymmetry of
CDM particles and antiparticles, (ii) the CDM annihilation cross-section
is greater than 10 pb so annihilation is complete (Scenario II), and
(iii) $m_d$ is not much greater than $5 m_p$. If
$m_d < \frac{1}{2} M_Z$, the measured $Z$ width rules out a 
Standard Model strength coupling of CDM to $Z$.
The nucleus recoil experiments have also ruled out
a Standard Model strength coupling of CDM to $Z$ with $m_d$ in the
range $10$ GeV to $10^3$ GeV. For $m_d > 10^3$ GeV and Standard Model
CDM-$Z$ coupling, Scenario I can not be reached.
So coupling to a new $Z'$ is favored. The large annihilation cross-section
needed, and the nucleus recoil experiments,
suggest that the $Z'$ does not couple to quarks if $m_d > 10$ GeV,
else there is fine tuning in resonant annihilation. So there is an
extra incentive to search for nucleus recoil for $1.5 \textrm{ GeV} < m_d < 10$ GeV.
Perhaps $Z'$ does not couple to quarks:
$Z'$ has not been seen at the Tevatron or the LHC, so if it exists,
its mass is high, and/or it does not couple, or couples weakly, to quarks.
Therefore there is an incentive to search for $Z'$ at a future $e^- e^+$ linear collider.
We would like to measure $e^-$ recoils in CDM-$e^-$ elastic scattering. 
The state-of-the art CCD detectors
have an electron energy threshold of 1.2 eV with a root-mean-square noise
of 0.2 electrons \cite{Skipper}, so \textit{single electron detection} is possible!
Measurement of the tail of the $e^-$ recoil distribution
may be feasible depending on the backgrounds.


\begin{thebibliography}{7}
\bibitem{PDG2014}
K.A. Olive et al. (Particle Data Group). Chin. Phys. C, 2014, \textbf{38}(9): 090001.
%J. Beringer et al. (Particle Data Group), Phys. Rev. D \textbf{86},
%010001 (2012).
%\bibitem{BH}
%The equations in Section \ref{scenarios} are derived
%in many books on Thermal Physics, including 
%B. Hoeneisen, ``Thermal Physics", Mellen Research University Press (1993).
\bibitem{gal}
B. Hoeneisen, arXiv:astro-ph/0009071 (2000).
\bibitem{cdm_LHC}
P. Calfayan (representing ATLAS and CMS), Rencontres de Moriond 2014.
Sarah Alam Malik (representing CMS), arXiv:1206.0753 (2012).
\bibitem{LUX}
D.S. Akerib et al. (LUX), arXiv:1310.8214v2 (2014).
\bibitem{mass}
B. Hoeneisen, arXiv:hep-th/0609080 (2006).
\bibitem{Branco}
Gustavo C. Branco, Lu\'{\i}s Lavoura, Jo\~{a}o Paulo Silva, ``CP violation",
Clarendon Press, 1999.
\bibitem{Illiopoulos}
J. Illiopoulos, Proceedings, Rencontres de Moriond 2014.
\bibitem{nuSM}
Laurent Canetti, Marco Drewes, Tibor Frossard, Mikhail Shaposhnikov,
Phys.Rev. D87 (2013) 9, 093006 (2012).
\bibitem{Skipper}
Guillermo Fernandez Moroni, Juan Estrada, Eduardo E. Paolini, 
Gustavo Cancelo, Stephen E. Holland, H. Thomas Diehl,
arXiv:1106.1839 (2011).
\end{thebibliography}
\end{document}